\documentclass[aps,twocolumn,showpacs,amsmath,amssymb,prc]{revtex4-2}
\usepackage{graphicx,here}
\usepackage{enumitem}
\usepackage{multirow}
\usepackage{tikz}
\usepackage{epstopdf}
\usepackage[percent]{overpic}
\usepackage{scrextend}
\usepackage{xcolor,xspace}
\usepackage[ulem=normalem]{changes}
\usepackage{hyperref}
\hypersetup{colorlinks=True,urlcolor=blue,linkcolor=blue,citecolor=blue,filecolor=black}

\colorlet{Changes@Color}{red}
\usepackage{dcolumn,color,footnote,bm,braket}
\usepackage{url,longtable,tabularx}
\usepackage{threeparttable}

\usepackage{fancybox}
\usetikzlibrary{matrix}

\newcommand\+{\dagger}
\newcommand\nb{N_{\rm B}}
\newcommand\pt{(p,t)}
\newcommand\tp{(t,p)}

\newcommand{\beonp}[4]{B(E3; #1^-_{#2} \!\to\! #3^+_{#4})}

\newcommand{\bej}[4]{#1^+_{#2} \!\to\! #3^+_{#4}}
\newcommand{\sigpt}[2]{I^{\pt}(0^+_{#1} \!\to\! 0^+_{#2})}
\newcommand{\sigtp}[2]{I^{\tp}(0^+_{#1} \!\to\! 0^+_{#2})}
\newcommand{\mpts}[2]{M^{\pt}_{s}(0^+_{#1} \!\to\! 0^+_{#2})}
\newcommand{\mptf}[2]{M^{\pt}_{sff}(0^+_{#1} \!\to\! 0^+_{#2})}
\newcommand{\mtps}[2]{M^{\tp}_{s}(0^+_{#1} \!\to\! 0^+_{#2})}
\newcommand{\mtpf}[2]{M^{\tp}_{sff}(0^+_{#1} \!\to\! 0^+_{#2})}

\newcommand{\ptra}[2]{^{#1}{\rm Ra}(p,t)^{#2}{\rm Ra}}
\newcommand{\tpra}[2]{^{#1}{\rm Ra}(t,p)^{#2}{\rm Ra}}
\newcommand{\ptth}[2]{^{#1}{\rm Th}(p,t)^{#2}{\rm Th}}
\newcommand{\tpth}[2]{^{#1}{\rm Th}(t,p)^{#2}{\rm Th}}

\newcommand{\orme}[1]{\braket{3^-_1\|\hat T(E3)\|0^+_{#1}}}
\newcommand{\ezero}{E_{\rm x}}
\newcommand{\stp}{R^{(t,p)}}
\newcommand{\spt}{R^{(p,t)}}

\newcommand{\nf}[3]{\braket{\hat n_f}_{{#1}^{#2}_{#3}}}

\begin{document}

\title{Octupole correlations in the
excited $0^+$ states and two-neutron
transfers in actinide nuclei}

\author{Kosuke Nomura}
\email{nomura@sci.hokudai.ac.jp}
\affiliation{Department of Physics, 
Hokkaido University, Sapporo 060-0810, Japan}
\affiliation{Nuclear Reaction Data Center, 
Hokkaido University, Sapporo 060-0810, Japan}

\date{\today}

\begin{abstract}
Impacts of octupole correlations
on low-energy excited $0^+$ states
and two-neutron transfer
intensities in Ra and Th isotopes
are analyzed within
the interacting boson model,
using the Hamiltonian parameters
that are obtained from microscopic
nuclear density functional theory
calculations.
Incorporating the octupole
degree of freedom in the model
is shown to give an overall
reasonable description
of the ground-state-band
energy levels and to lower the
$0^+_2$ ones significantly.
The calculation yields a number
of excited $0^+$ states,
most of which are formed
by the configurations
of more than one octupole boson.
Octupole correlations make
significant contributions to
the $0^+_1 \to 0^+_2$
$(p,t)$ and $(t,p)$
transfer intensities in those
nuclei in which the octupole
collectivity is most enhanced.
Behaviors of the predicted
transfer intensities with the increasing
neutron number along the
isotopic chain reflect nuclear
structural evolution.
\end{abstract}

\maketitle

\section{Introduction}

The intrinsic shape of the atomic nucleus is
characterized mostly by the quadrupole
deformations \cite{BM}.
A higher-order deformation of octupole
type comes into play in low-lying states
in those nuclei with
specific neutron $N$
and/or proton $Z$ numbers, e.g.,
34, 56, 88, 134, for which
a $\Delta j=\Delta l=3$
coupling of the single-particle
orbits is possible
\cite{ahmad1993,butler1996,butler2016,butler2020b,butler2024}.
Evidence for
the permanent octupole
deformation was found
in a few radioactive nuclei,
e.g., $^{220}$Rn and $^{224}$Ra
\cite{gaffney2013},
which exhibit a negative-parity,
$K^{\pi}=0^-_1$, band close
in energy to the ground-state,
$K^{\pi}=0^+_1$, band;
they form an approximate
alternating-parity band that
is connected by strong electric
dipole ($E1$) and octupole ($E3$)
transitions.
While the static octupole
shape rarely occurs in real nuclei,
the octupole degree of freedom
and its coupling to the quadrupole
degree of freedom are important building
blocks to describe low-energy
nuclear structure.

One of the peculiar features
of the nuclear structure is
the presence of low-energy
excited $0^+$ states
and corresponding bands,
which have been observed
in many nuclei.
These excited $0^+$ energy levels
often exhibit a parabolic behavior
with $N$ in a given isotopic chain,
being particularly low near the middle
of the major shell.
These characteristics point to
the coexistence of multiple intrinsic shapes
\cite{heyde1983,wood1992,heyde2011,garrett2022,leoni2024},
considered nowadays to be
a universal phenomenon that is
found in a wide range of the nuclear chart.
In the rare-earth region,
the lowering pattern of the $0^+$
energy levels has also been identified
as a signature of the quantum
phase transitions (QPTs)
in nuclear shapes
\cite{carr-book,casten2009,cejnar2010,fortunato2021}.
An alternative interpretation has been
provided in terms of the
pairing vibrations
\cite{bes1966,bes1972,broglia1973,
prochniak1999,
brink-broglia,
srebrny2006,
prochniak2007,
xiang2020,nomura2020pv}.
Nuclei in rare-earth and
actinide regions with
$(N,Z)\approx(88,56)$
and $(134,88)$, respectively,
are characterized by the
pronounced octupole collectivity.
The low-energy $0^+$ states
observed in these regions
have been suggested to be
of two-octupole-phonon nature
\cite{butler2016}.
Related experiments were
performed, e.g., by
means of the two-neutron transfer reactions
\cite{BM,brink-broglia}
for actinide
\cite{wirth2004,levon2009,levon2013,spieker2013,spieker2018}
and rare-earth
\cite{zamfir2002,meyer2006-pt,
pascu2009,pascu2010,bucurescu2019,
levon2020-158Gd-pt}
nuclei.
The interpretation of the
low-energy $0^+$ states and the
corresponding $K^{\pi}=0^+$ bands
remains an open problem,
and it would be of interest to
give an appropriate theoretical description
of these phenomena that is based on a certain microscopic
nuclear structure.

The present study aims to address
the octupole correlation effects
on the ground and excited $0^+$ states
and two-neutron transfer reactions
in actinides, using the
interacting boson model (IBM) \cite{IBM}.
The IBM comprises as basic building blocks
correlated monopole and quadrupole
pairs of valence nucleons,
which are approximated by $s$ and $d$
bosons, respectively.
It has been successful in describing
low-energy quadrupole collective states
in medium-heavy and heavy nuclei,
and was also applied
to compute $\pt$ and $\tp$
transfer intensities as signatures
of the QPTs
\cite{arima1977trf,fossion2007,zhang2017,
nomura2019trf,garciaramos2020}.
In a recent study
\cite{nomura2026trf-1},
octupole correlation effects on the
excited $0^+$ states and two-neutron
transfer reactions in Nd, Sm, and Gd isotopes
near the shape transition point,
$N\approx90$,
were studied using the IBM that
includes the octupole, $f$, boson
degree of freedom,
using the Hamiltonian parameters
derived by means of the self-consistent
mean-field (SCMF) method in the
framework of the
nuclear density functional theory (DFT)
\cite{RS,bender2003,vretenar2005,robledo2019}.
This version of the IBM,
denoted hereafter mapped IBM,
has been successfully used
in nuclear structure studies
(see Ref.~\cite{nomura2025rev} for a
review and references therein).
The $sdf$-IBM calculations
in Ref.~\cite{nomura2026trf-1}
produced a
number of excited $0^+$ states that
consist of the configurations of
more than one $f$ boson,
and produced QPT-like behaviors of the
$0^+ \to 0^+$
$\pt$ and $\tp$ transfer intensities
as functions of $N$.

The present study extends the analysis
of Ref.~\cite{nomura2026trf-1} to
the even-even actinide, Ra and Th nuclei,
in which enhanced
octupole collectivity is expected.
Previous DFT-mapped IBM studies
\cite{nomura2013oct,nomura2014,nomura2020oct}
suggested an occurrence of
octupole shape phase transitions
in Th and Ra isotopes,
that is, transitions from
nearly spherical to stable
octupole and to
octupole-soft shapes.
In these studies,
while the main focus was placed on
the properties of the yrast states,
possible double-octupole-phonon
nature of the $0^+_2$
states was also pointed out in
a few selected nuclei.
The scope of the present study is
to provide a more
systematic investigation of
the relevance of the octupole
collectivity in the $0^+$ states in
Ra and Th.
As compared with
the rare-earth nuclei,
experimental information
of the two-neutron transfers
in actinide region is rather scarce.
The IBM calculations
with the $s$, $d$, $f$, and $p$
(or dipole) boson degrees of freedom
were made to interpret
$\pt$ data for a few nuclei,
$^{228}$Th \cite{wirth2004,levon2013},
$^{230}$Th \cite{wirth2004,levon2009},
$^{232}$U \cite{wirth2004},
and $^{240}$Pu \cite{spieker2013,spieker2018},
in which parameters of the $spdf$-IBM
were fitted to experimental
low-energy spectra.
Predictions of
the $\pt$ and $\tp$ transfer intensities
are here made
within the mapped IBM, which is guided
by the self-consistent calculations within the
nuclear DFT.
In particular, spectroscopic properties of
the Ra and Th nuclei considered in the
present work were studied
in the previous mapped IBM calculation
in Ref.~\cite{nomura2020oct},
with the microscopic
input being provided by
the Hartree-Fock-Bogoliubov
(HFB) \cite{robledo2019}
self-consistent
calculations using the D1M
parametrization \cite{D1M}
of the Gogny force \cite{Gogny}.
The derived values of the IBM parameters
in Ref.~\cite{nomura2020oct}
are employed in the present work.

In Sec.~\ref{sec:model},
a brief description of
the theoretical framework to
compute the two-nucleon transfers
is given.
Calculated low-energy
spectra, transition properties,
and $\pt$ and $\tp$
transfer intensities
are presented in Sec.~\ref{sec:results}.
A summary of main results is
provided in Sec.~\ref{sec:summary}.

\section{Theoretical procedure\label{sec:model}}

The collective quadrupole and octupole
states are described in terms
of the $s$, $d$, and $f$ boson
degrees of freedom
\cite{scholten1978,engel1985,engel1987,barfield1988,IBM},
which represent collective $0^+$,
$2^+$, and $3^-$ pairs of valence
nucleons, respectively \cite{OAIT,OAI,IBM}.
The present version of the IBM
does not distinguish
like-proton from like-neutron bosons,
and is referred to as $sdf$-IBM-1,
which is hereafter denoted $sdf$-IBM
unless otherwise specified.
The number of bosons, denoted
$\nb$, is equal to
that of pairs of
valence nucleons, and is constant
for a given nucleus.

The $sdf$-IBM Hamiltonian adopted
in this study is of the form
\begin{align}
\label{eq:ham}
\hat H_{\rm B}= 
\epsilon_d\hat n_{d} 
+ \epsilon_{f}\hat{n}_{f} 
+ \kappa_{2}\hat{Q}_{2}\cdot\hat{Q}_{2}
+ \kappa_{3}\hat{Q}_{3}\cdot\hat{Q}_{3}
+ \rho\hat{L}_d\cdot\hat{L}_d \; .
\end{align}
The first and second terms, respectively,
denote the $d$-boson and $f$-boson
number operators, defined as
$\hat n_{d}=d^{\+}\cdot\tilde d$
and $\hat n_f=f^{\+}\cdot\tilde f$,
with $\epsilon_d$ and
$\epsilon_f$ being the single $d$-boson
and $f$-boson energies relative to that
of $s$ bosons.
Note that $\tilde d_{\mu}=(-1)^{-\mu}d_{-\mu}$
and $\tilde f_{\mu}=(-1)^{3-\mu}f_{-\mu}$.
The third and fourth terms represent
the quadrupole-quadrupole and octupole-octupole
interactions with strengths $\kappa_2$
and $\kappa_3$, respectively.
The last term concerns the corrections
to the rotational bands.
The corresponding boson operators
are defined as
\begin{subequations}
 \begin{align}
\label{eq:q2}
& \hat Q_{2}=s^{\dagger}\times\tilde d
+d^{\dagger}\times\tilde s
+\chi_{d}(d^{\dagger}\times\tilde d)^{(2)}
+\chi_{f}(f^{\dagger}\times\tilde f)^{(2)} \\
\label{eq:q3}
&\hat{Q}_{3}=
s^{\+}\times\tilde{f}
+f^{\+}\times\tilde{s}
+\chi_{3}(d^{\+}\times\tilde{f}
+f^{\+}\times\tilde{d})^{(3)}
\\
\label{eq:l}
& \hat{L}_d=
\sqrt{10}(d^{\+}\times\tilde{d})^{(1)}
\; .
\end{align}
\end{subequations}
The use of the Hamiltonian of
the form \eqref{eq:ham} is inspired
by the earlier microscopic IBM study
of Ref.~\cite{barfield1988},
in which an $sdf$-IBM-1 Hamiltonian
similar to \eqref{eq:ham} was
derived from a proton-neutron
octupole-octupole boson interaction
by mapping the totally symmetric
state in the proton-neutron $sdf$-IBM onto
the corresponding IBM-1 state.
The same form of the $sdf$-IBM Hamiltonian
as in \eqref{eq:ham} was used in
Ref.~\cite{nomura2015}
for studying quadrupole-octupole collective
states in Sm and Gd isotopes.

The parameters of $\hat H_{\rm B}$ are
derived by associating
the $sdf$-IBM Hamiltonian
to the SCMF potential energy surface (PES)
obtained from the axial quadrupole,
$\beta_{20}$,
and octupole, $\beta_{30}$,
constrained Gogny-HFB calculations,
with $\beta_{20}$ and $\beta_{30}$
representing intrinsic deformations
related to the axial
quadrupole and octupole moments,
respectively.
The intrinsic properties of a given
nucleus is inferred in the $sdf$-IBM
from the consideration of the boson
coherent state,
\begin{equation}
 \ket{\phi}=(\nb!)^{-1/2}(\lambda^{\+})^{\nb}\ket{0}
\end{equation}
with
\begin{equation}
\lambda^{\+}=s^{\+}+\tilde\beta_{20}d^{\+}_{0}+\tilde\beta_{30}f^{\+}_0
\; .
\end{equation}
$\tilde\beta_{\lambda0}$ are
bosonic quadrupole and octupole
deformations, and are assumed to be
proportional to the fermionic
intrinsic deformations
as $\tilde\beta_{\lambda0}=C_{\lambda}\beta_{\lambda0}$,
with $C_{\lambda}$ being the constants
of proportionality
\cite{nomura2008,nomura2013oct}.
Taking the expectation value of
$\hat H_{\rm B}$ in the coherent state
$\ket{\phi}$ gives rise to the
energy surface in the $sdf$-boson
system.
The $sdf$-IBM PES,
$E_{\rm IBM}(\beta_{20},\beta_{30})$,
is then associated with that calculated within
the constrained Gogny-HFB method,
$E_{\rm HFB}(\beta_{20},\beta_{30})$,
so that an approximate relation
\begin{equation}
E_{\rm IBM}(\beta_{20},\beta_{30})
\approx
E_{\rm HFB}(\beta_{20},\beta_{30})
\end{equation}
should hold in the vicinity
of the equilibrium minimum of the PES.
The strength $\rho$ for the
$\hat L_d \cdot \hat L_d$ term is
obtained independently of the above
PES-mapping procedure
so that \cite{nomura2011rot}
the bosonic
cranking moment of inertia
\cite{schaaser1986}
at the equilibrium point along the
$\beta_{30}=0$ axis
should be equal to the HFB counterpart
computed by the Thouless-Valatin
formula \cite{TV}.
The details of the $sdf$-IBM
mapping procedure and the adopted
values of the parameters are found in Ref.~\cite{nomura2020oct}.
Note that the $\epsilon_f$ energies used
in Ref.~\cite{nomura2020oct}
are here modified, because
the $\hat L_{f}\cdot\hat L_{f}$ term,
with $\hat L_f=\sqrt{28}(f^{\+}\times\tilde f)^{(1)}$,
was contained in the Hamiltonian in
Ref.~\cite{nomura2020oct},
but is not in the present work.
The mapped $sdf$-IBM Hamiltonian
is diagonalized in the boson
$m$-scheme basis \cite{nomura2022octcm},
providing excitation energies
and wave functions.

The $\pt$ and $\tp$ transfer
strengths are computed as
\begin{align}
\label{eq:sigpt}
& \sigpt i f
=|\braket{N-2,0^+_f| \hat P^{\pt} | N,0^+_i}|^2
\\
\label{eq:sigtp}
& \sigtp i f
=|\braket{N+2,0^+_f | \hat P^{\tp} | N,0^+_i}|^2
\; .
\end{align}
$\ket{N,0^+_{i}}$
and $\ket{N\mp2,0^+_{f}}$ stand for
the $sdf$-IBM wave functions of the
$0^+$ states in an initial
and a final nuclei
with $N$ and $N\mp2$ in the $\pt$
and $\tp$ reactions, respectively.
$\hat P^{\pt}$ and $\hat P^{\tp}$ are
bosonic $\pt$ and $\tp$ transfer operators
and are expressed as
\begin{eqnarray}
\label{eq:pt}
\hat P^{\pt}
=
c_{1}
\left[
\hat A(\Omega_{\nu},N_{\nu})
\tilde s
 + c_2
\hat n_{f}
\tilde s
\right]
\end{eqnarray}
and
\begin{eqnarray}
\label{eq:tp}
\hat P^{\tp}
=
c_{1}
\left[
s^{\+}
\hat A(\Omega_{\nu},N_{\nu})
 + c_2
s^{\+}
\hat n_{f}
\right] \; ,
\end{eqnarray}
respectively,
where $\hat P^{\tp}
=\left(\hat P^{\pt}\right)^{\+}$,
and $c_1$ and $c_2$
are parameters.
The operator
$\hat A(\Omega_\nu,N_\nu)$
is defined as \cite{arima1977trf}
\begin{align}
\label{eq:fac}
 \hat A(\Omega_\nu,N_\nu)
=
\left(
\Omega_\nu-N_\nu-\frac{N_\nu}{N_{\rm B}}\hat n_{d}
\right)^{1/2}
\left(
\frac{N_{\nu}+1}{\nb+1}
\right)^{1/2}
\; ,
\end{align}
where $\Omega_\nu$ denotes
the degeneracy
of the neutron pairs in a given major
shell, and is equal to
$(184-126)/2=29$
for the actinide nuclei under consideration.
$N_{\nu}$ stands for the
number of like-neutron
bosons, i.e.,
$N_{\nu}=1$, 2, $\ldots$, 12
for those even-even nuclei with
$N=130$, 132, $\ldots$, 150, respectively.
As in literature \cite{IBM},
the operator $\hat n_{d}$
in Eq.~(\ref{eq:fac}) is
replaced with its expectation
value in the ground state
of the initial nucleus,
i.e.,
$\braket{0^+_{1,i}|\hat n_{d}|0^+_{1,i}}$.

The operators similar to those
in \eqref{eq:pt} and \eqref{eq:tp}
were frequently used in the
phenomenological $spdf$-IBM
calculations for two-neutron transfers
in actinides
\cite{wirth2004,levon2009,levon2013,spieker2013,spieker2018},
in which in addition to the
$s(^{\+})\hat n_f$ term the term
that involves $p$ boson operator,
$s(^{\+})\hat n_p$, was considered.
It was also shown that
for the descriptions of the
two-nucleon transfers contributions
from the negative-parity bosons were
essential.
The first terms in the operators
in \eqref{eq:pt} and \eqref{eq:tp},
$\hat A(\Omega_{\nu},N_{\nu})s(^{\+})$,
in fact, arise from the seniority
consideration within the $sd$-IBM \cite{arima1977trf},
while the second terms, $s(^{\+})\hat n_f$,
are specific terms involved in more
general transfer operators in the $sdf$-IBM
and the addition of these terms
to $\hat A(\Omega_{\nu},N_{\nu})s(^{\+})$
may not be fully justified from
a microscopic point of view.
However, the use of these specific forms,
\eqref{eq:pt} and \eqref{eq:tp},
are just adequate for the present purpose,
which is to study the
impacts of the octupole degree of
freedom on the transfer strengths.

In addition, to analyze
the $f$-boson
effects on spectroscopic properties
quantitatively,
the $sd$-IBM mapping calculations are
also performed, with
the maximum number of $f$ bosons
being set to zero.
The corresponding transfer operators
in the $sd$-boson model thus
consist only of the
first terms in \eqref{eq:pt}
and \eqref{eq:tp}.

%-----------------------------------------------------------
%
%   \pi = + 1 yrast levels
%
%-----------------------------------------------------------
\begin{figure}[htb!]
\begin{center}
\includegraphics[width=\linewidth]{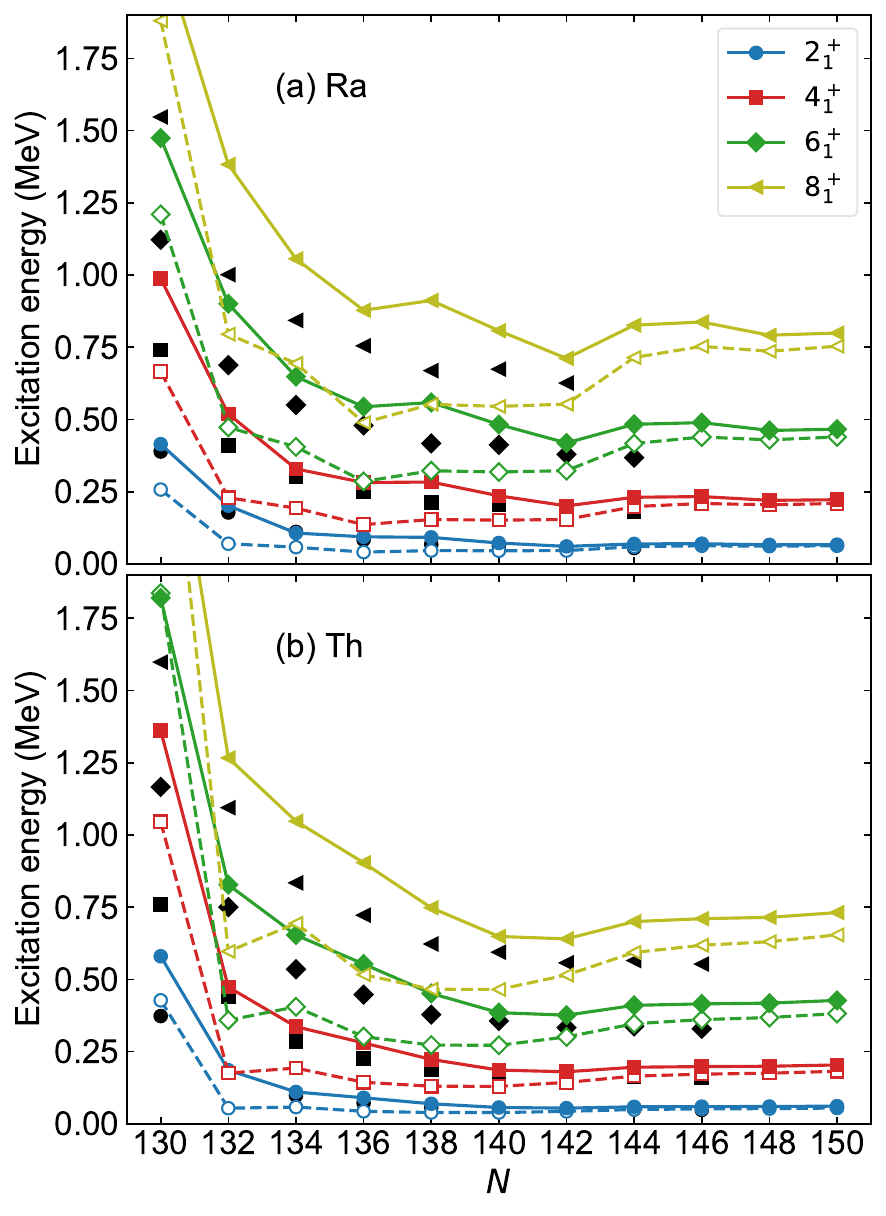}
\caption{Energy spectra of
even-spin positive-parity yrast states
in Ra and Th isotopes
calculated with the
mapped $sdf$-IBM and $sd$-IBM,
represented by
the solid and open symbols connected by
the solid and dashed lines, respectively.
The experimental data
are adopted from Ref.~\cite{data},
and are shown as
the black solid symbols.}
\label{fig:pos}
\end{center}
\end{figure}

%-----------------------------------------------------------
%
%   \pi = - 1 yrast levels
%
%-----------------------------------------------------------
\begin{figure}[htb!]
\begin{center}
\includegraphics[width=\linewidth]{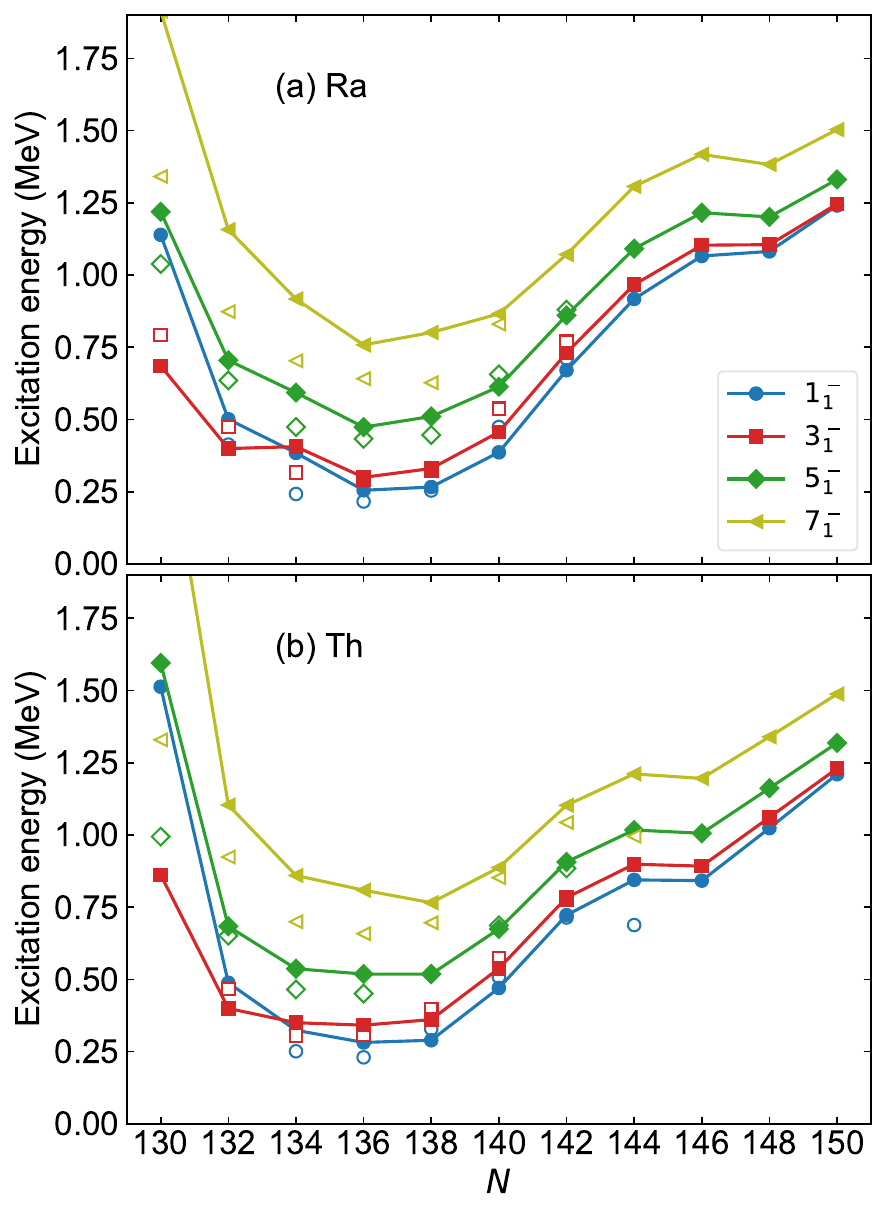}
\caption{Energy spectra for the odd-spin
negative-parity yrast states obtained
from the mapped $sdf$-IBM,
represented by the solid symbols
connected by the solid lines.
Experimental data, adopted
from Ref.~\cite{data}, are depicted
as open symbols.}
\label{fig:neg}
\end{center}
\end{figure}

%-----------------------------------------------------------
%
%   nonyrast
%
%-----------------------------------------------------------
\begin{figure}[htb!]
\begin{center}
\includegraphics[width=\linewidth]{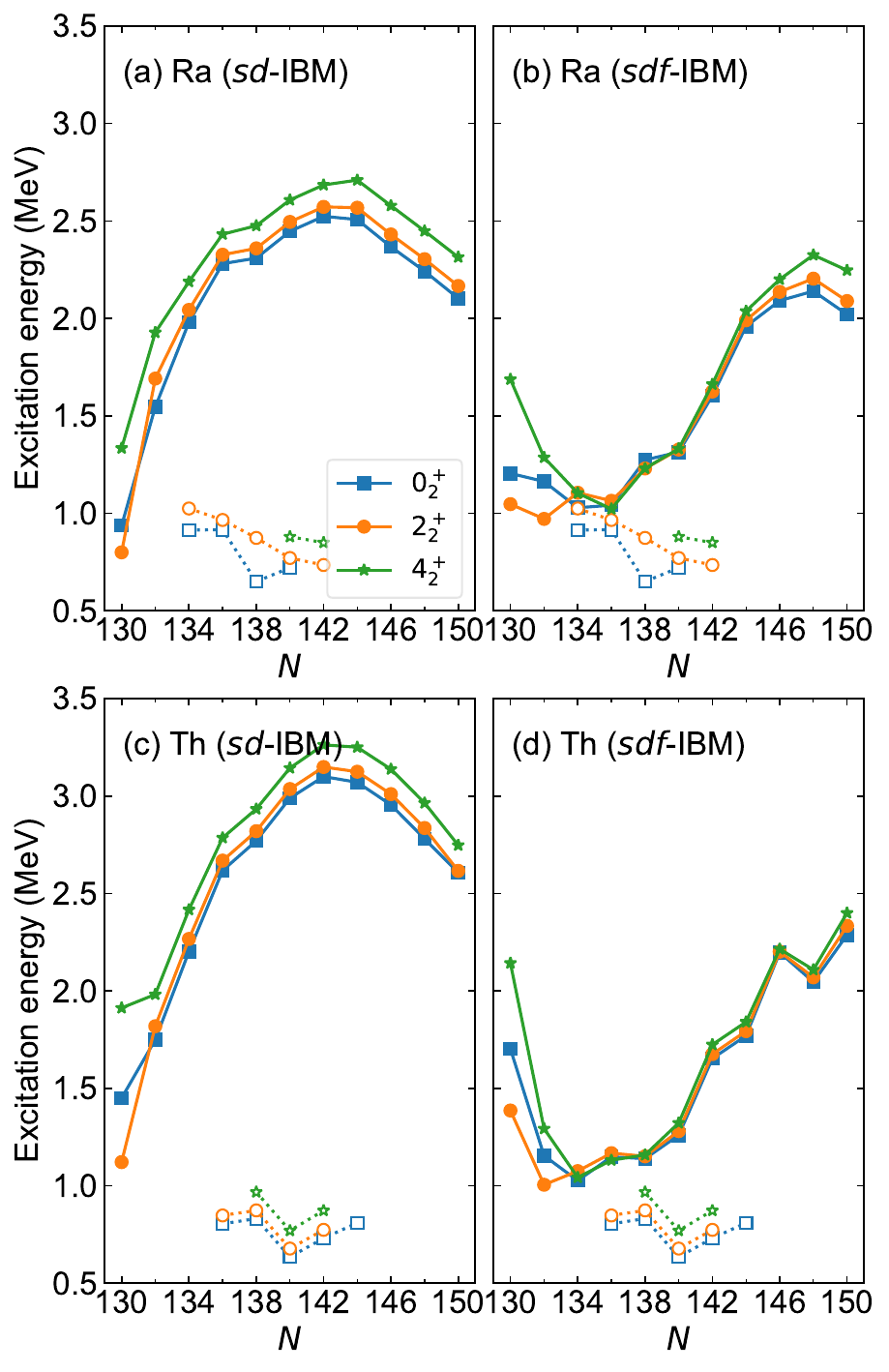}
\caption{Predicted
excitation energies of nonyrast states,
$0^+_2$, $2^+_2$, and $4^+_2$,
within the $sd$-IBM [(a) and (c)]
and $sdf$-IBM [(b) and (d)],
represented by the solid
symbols connected by the
solid lines.
The corresponding experimental
data, adopted from Ref.~\cite{data},
are depicted as open symbols
connected by dotted lines.}
\label{fig:ny}
\end{center}
\end{figure}

%-----------------------------------------------------------
%
%   f-boson num.
%
%-----------------------------------------------------------
\begin{figure*}[htb!]
\begin{center}
\includegraphics[width=.7\linewidth]{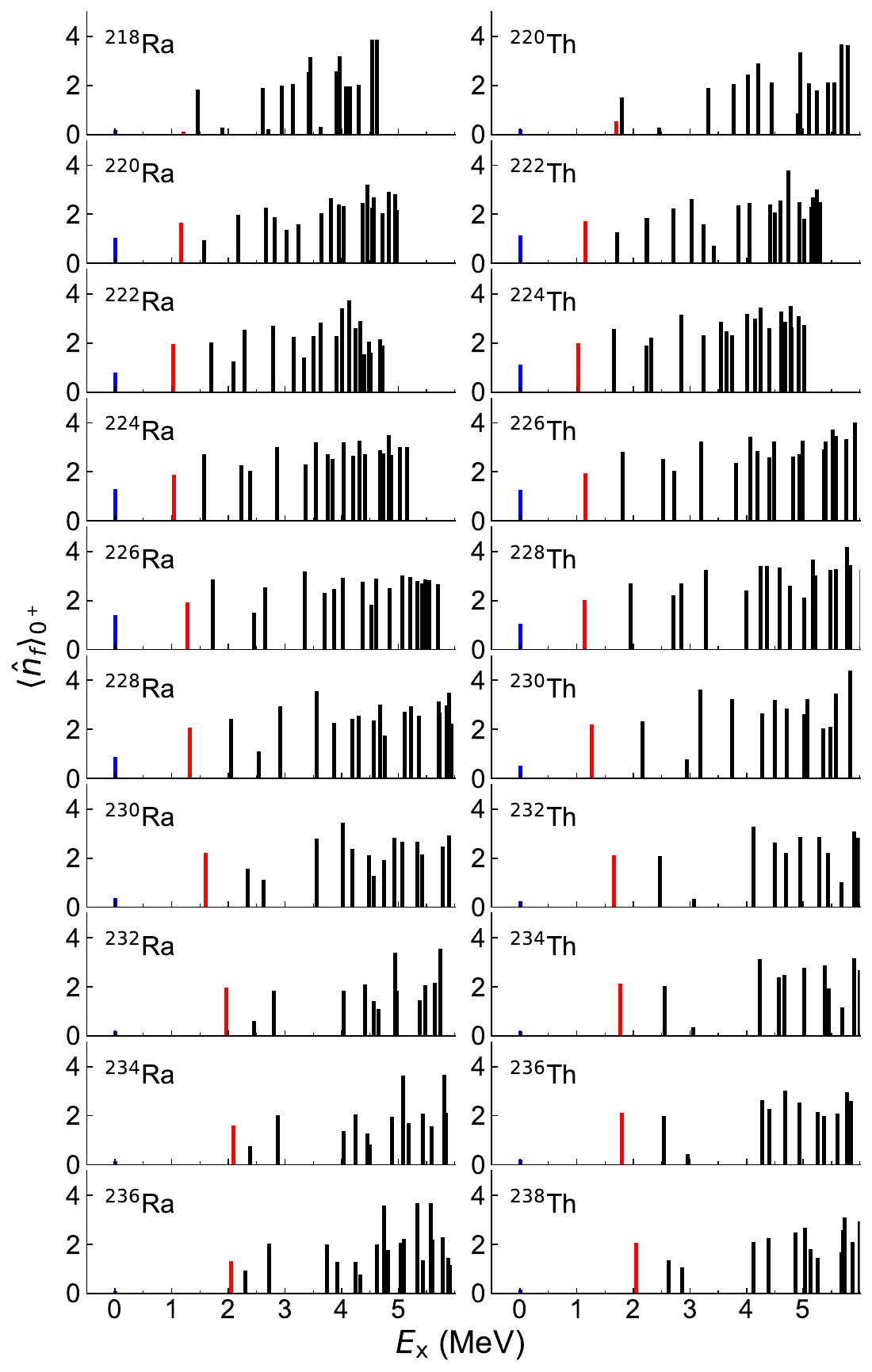}\\
\caption{Expectation values of
the $f$-boson
number operator, $\nf 0 + {n}$ ($n\geqslant 1$),
in the $0^+_n$ states in Ra and Th
isotopes, plotted against the
corresponding energies, $\ezero$.
The $\nf 0 + 1$ and $\nf 0 + 2$ values
are highlighted in blue and red,
respectively.}
\label{fig:nf}
\end{center}
\end{figure*}

%-----------------------------------------------------------
%
%   E3 r.m.e.
%
%-----------------------------------------------------------
\begin{figure}[htb!]
\begin{center}
\includegraphics[width=\linewidth]{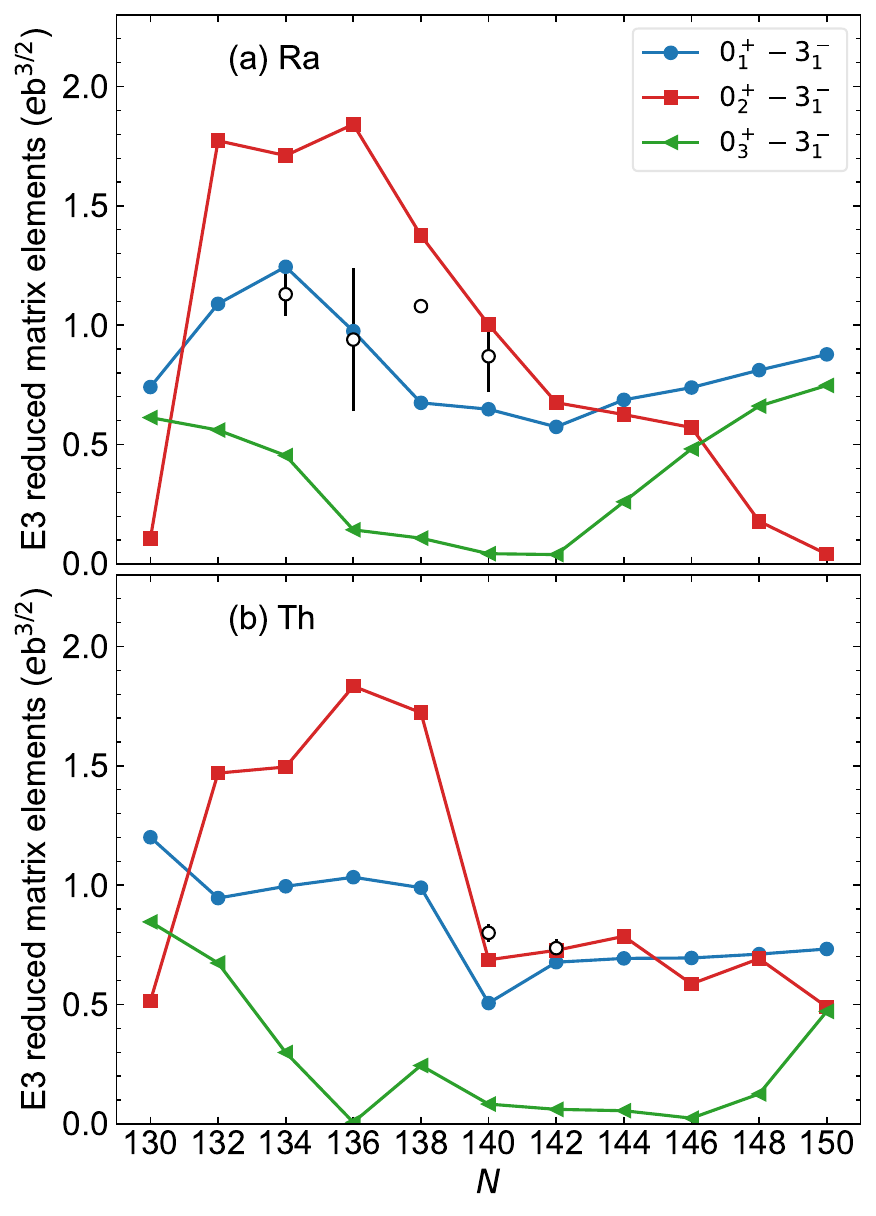}
\caption{Absolute values
of the calculated reduced matrix elements
$|\orme 1|$, $|\orme 2|$, and $|\orme 3|$
in $e$b$^{3/2}$ units.
Experimental $|\orme 1|$ values
for $^{222,228}$Ra \cite{butler2020a},
$^{224}$Ra \cite{gaffney2013},
$^{226}$Ra \cite{wollersheim93},
$^{230}$Th \cite{mcgowan1974},
and $^{232}$Th \cite{mcgowan1993}
are represented by open circles.}
\label{fig:rme}
\end{center}
\end{figure}

%-----------------------------------------------------------
%
%   (p,t) and (t,p) oct. Ra
%
%-----------------------------------------------------------
\begin{figure*}[htb!]
\begin{center}
\includegraphics[width=\linewidth]{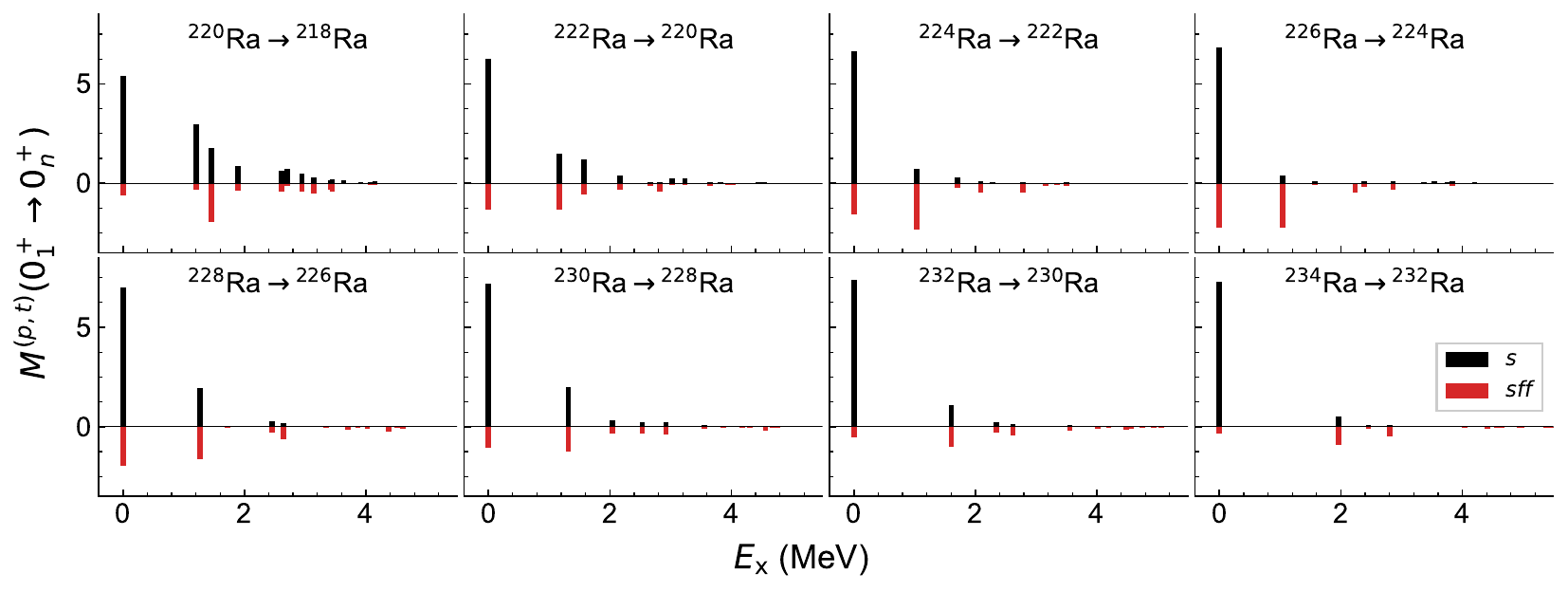}
\includegraphics[width=\linewidth]{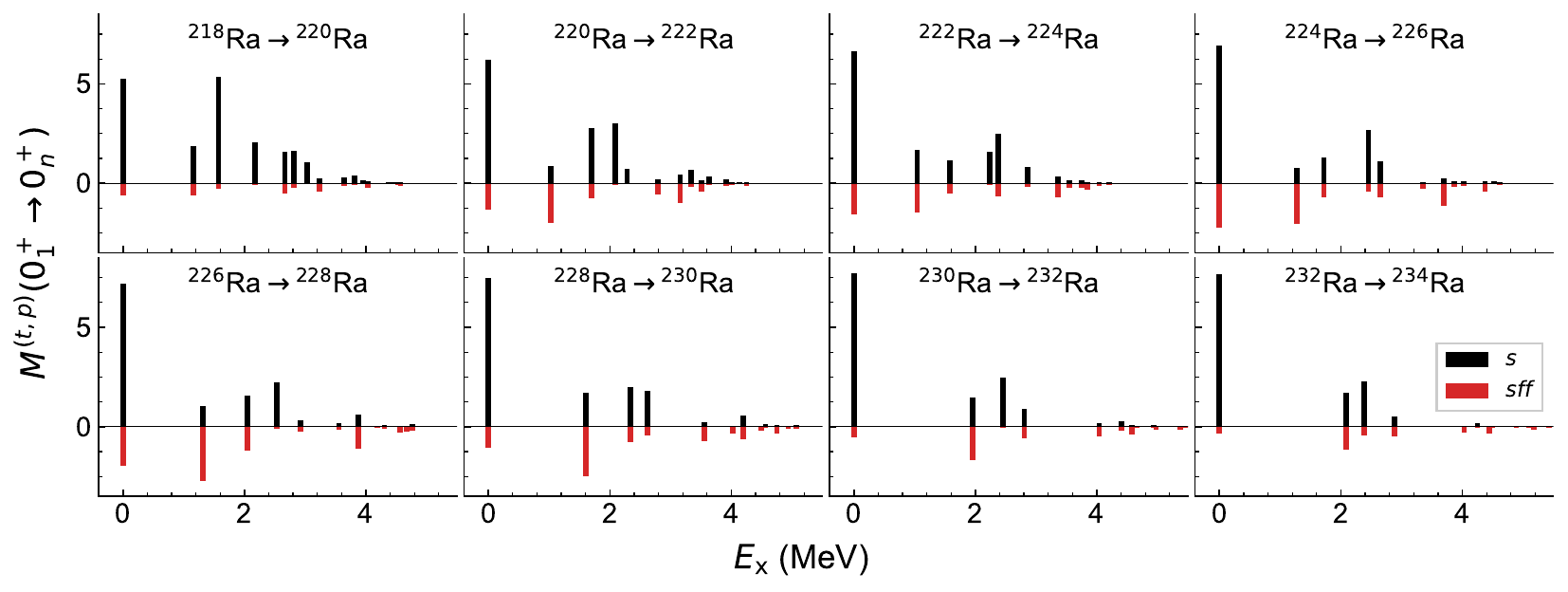}
\caption{Calculated matrix elements,
\eqref{eq:mpts}--\eqref{eq:mtpf}
with $n\geqslant 1$,
for the $\ptra {A+2} {A}$ (upper)
and $\tpra A {A+2}$ (lower)
reactions with $A=218-232$,
plotted against
the excitation energies $\ezero$
in the final nucleus.
In each plot, the absolute values of the
$s$-boson contributions,
$|\mpts 1 {n}|$ and $|\mtps 1 {n}|$
(denoted ``$s$''),
are shown, while the $f$-boson contributions
are shown with their sign being negative,
that is, $-|\mptf 1 {n}|$
and $-|\mtpf 1 {n}|$ (denoted ``$sff$'').
}
\label{fig:oct-trf-ra}
\end{center}
\end{figure*}

%-----------------------------------------------------------
%
%   (p,t) and (t,p) oct. Th
%
%-----------------------------------------------------------
\begin{figure*}[htb!]
\begin{center}
\includegraphics[width=\linewidth]{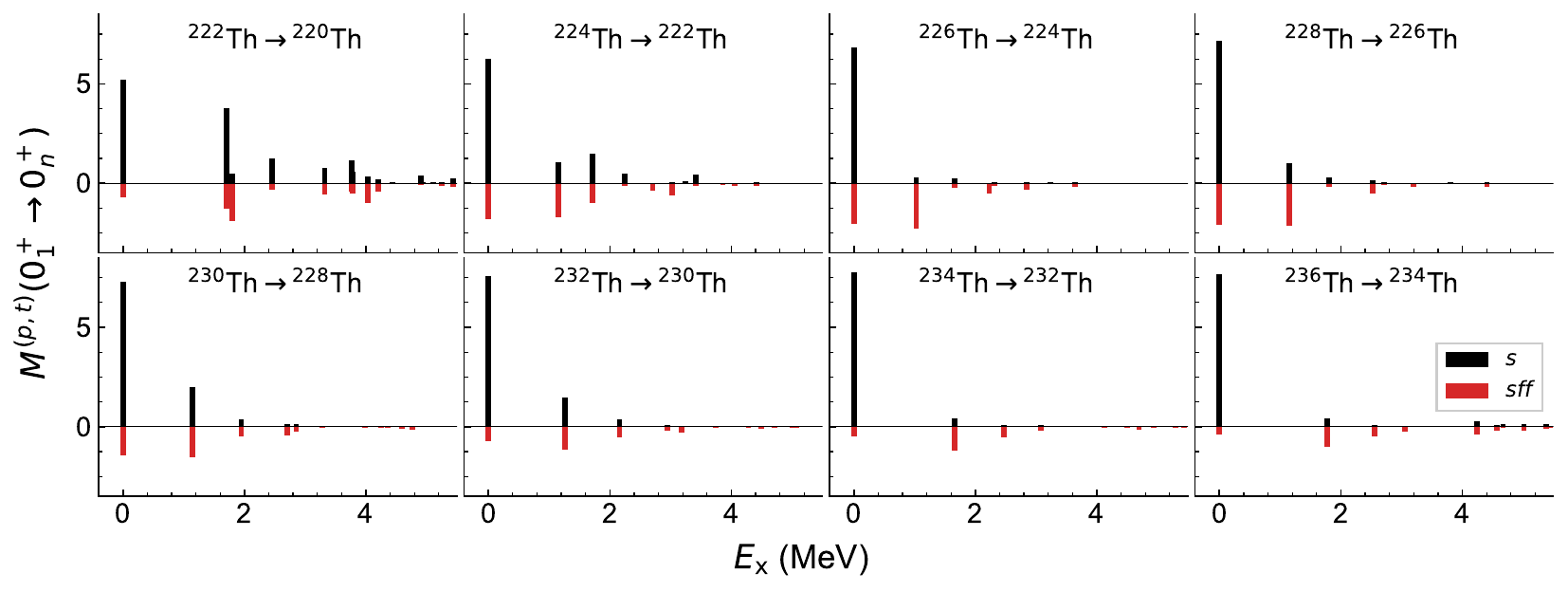}
\includegraphics[width=\linewidth]{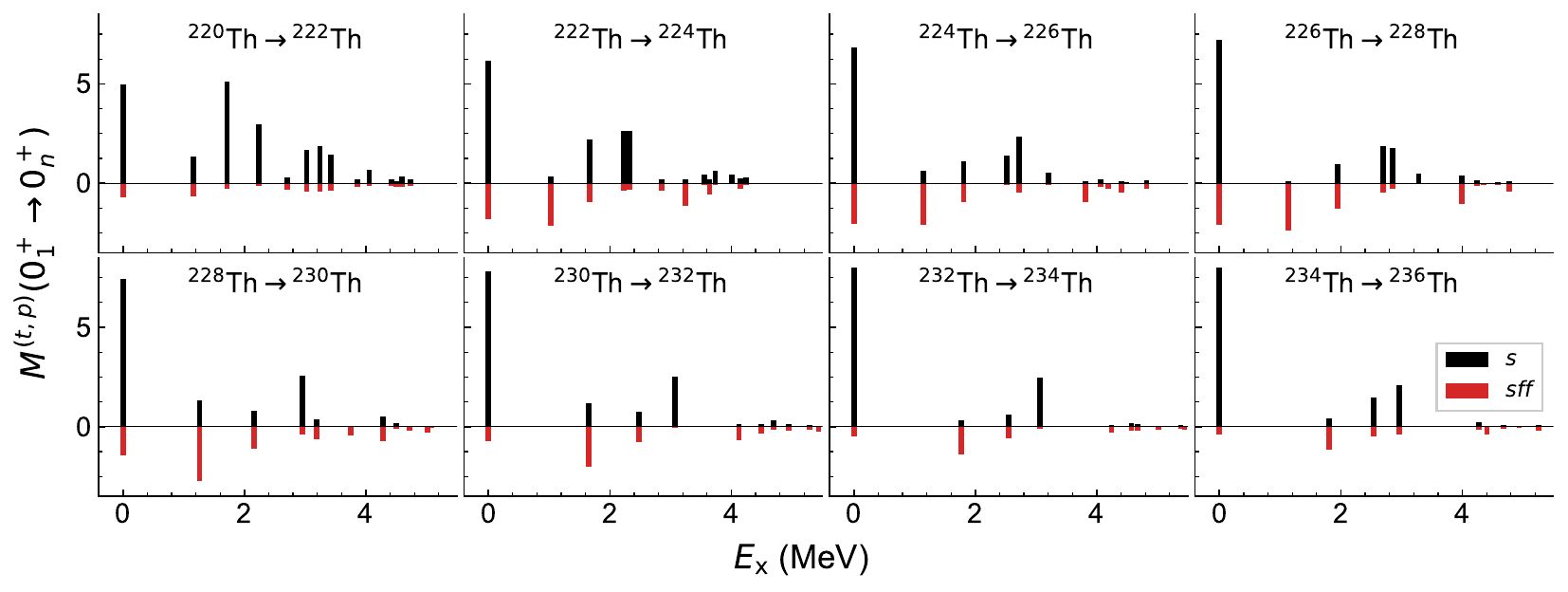}
\caption{As in Fig.~\ref{fig:oct-trf-ra}, but for the
$\ptth {A+2} {A}$ (upper) and $\tpth A {A+2}$ (lower)
reactions with $A=220-234$.}
\label{fig:oct-trf-th}
\end{center}
\end{figure*}

%-----------------------------------------------------------
%
%   Intensity, Ra
%
%-----------------------------------------------------------
\begin{figure*}[htb!]
\begin{center}
\includegraphics[width=.7\linewidth]{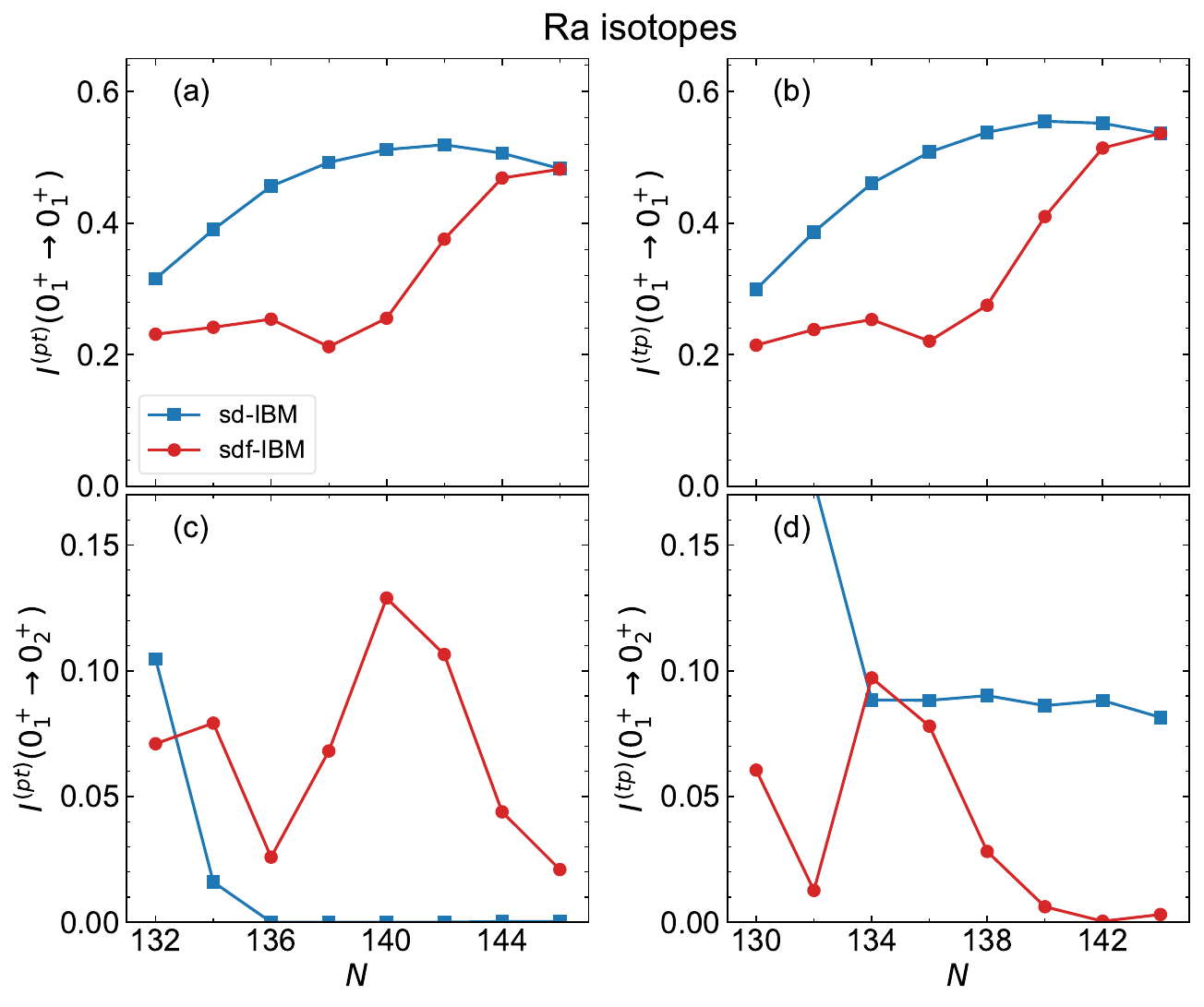}
\caption{$\pt$ and $\tp$ transfer
intensities in Ra isotopes depicted as functions
of $N$ of the initial nucleus.
Results obtained from the
mapped $sdf$-IBM are compared with those
from the $sd$-IBM.}
\label{fig:int-ra}
\end{center}
\end{figure*}

%-----------------------------------------------------------
%
%   Intensity, Th
%
%-----------------------------------------------------------
\begin{figure*}[htb!]
\begin{center}
\includegraphics[width=.7\linewidth]{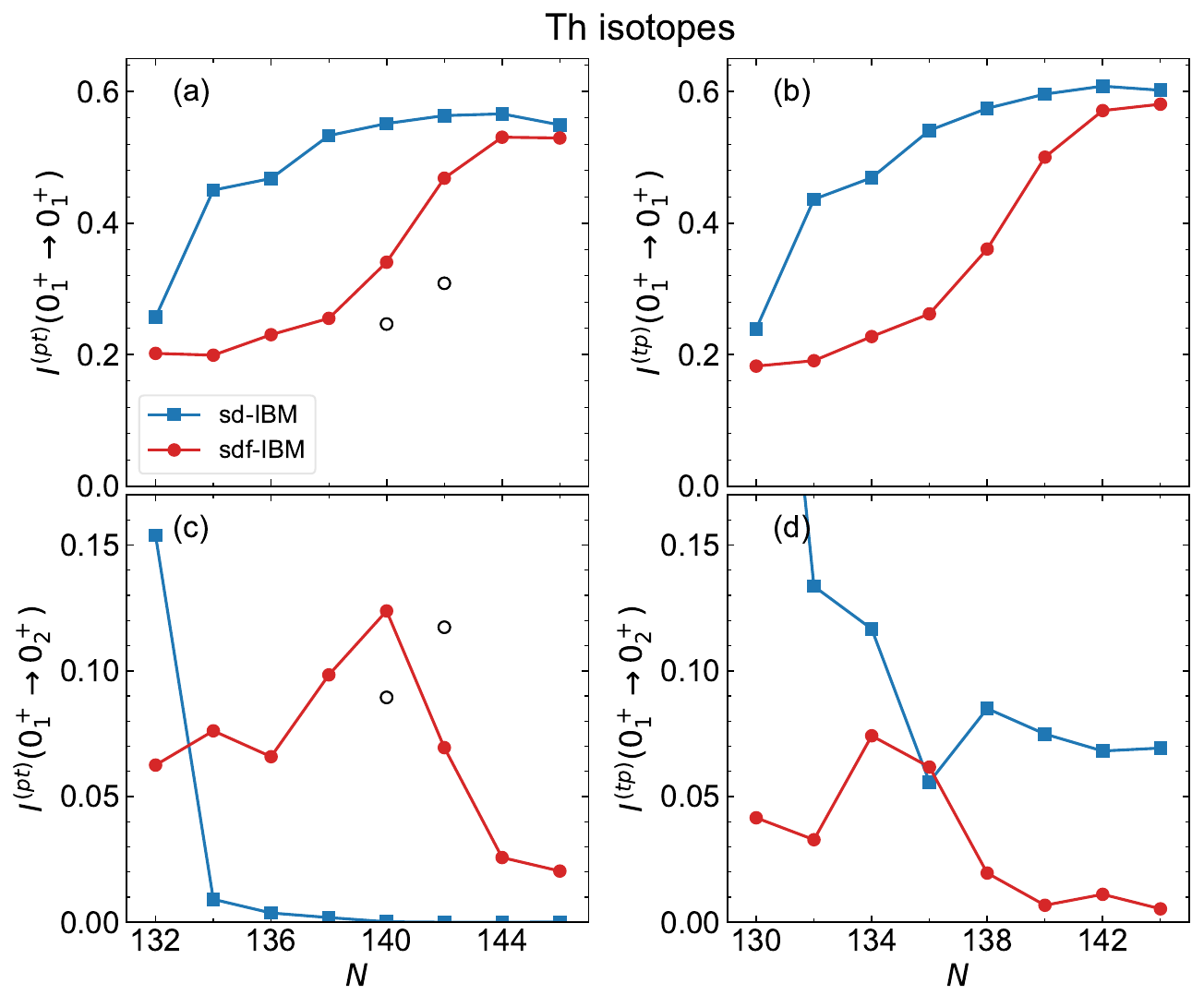}
\caption{As in Fig.~\ref{fig:int-ra},
but for the Th isotopes.
Experimental data for the $\ptth {230} {228}$
and $\ptth {232} {230}$,
adopted from Ref.~\cite{wirth2004},
are differential cross sections
in $\mu$b/sr units
measured at the laboratory angle
of $7.5^{\circ}$.}
\label{fig:int-th}
\end{center}
\end{figure*}

%-----------------------------------------------------------
%
%   runsum, Ra
%
%-----------------------------------------------------------
\begin{figure*}[htb!]
\begin{center}
\includegraphics[width=.7\linewidth]{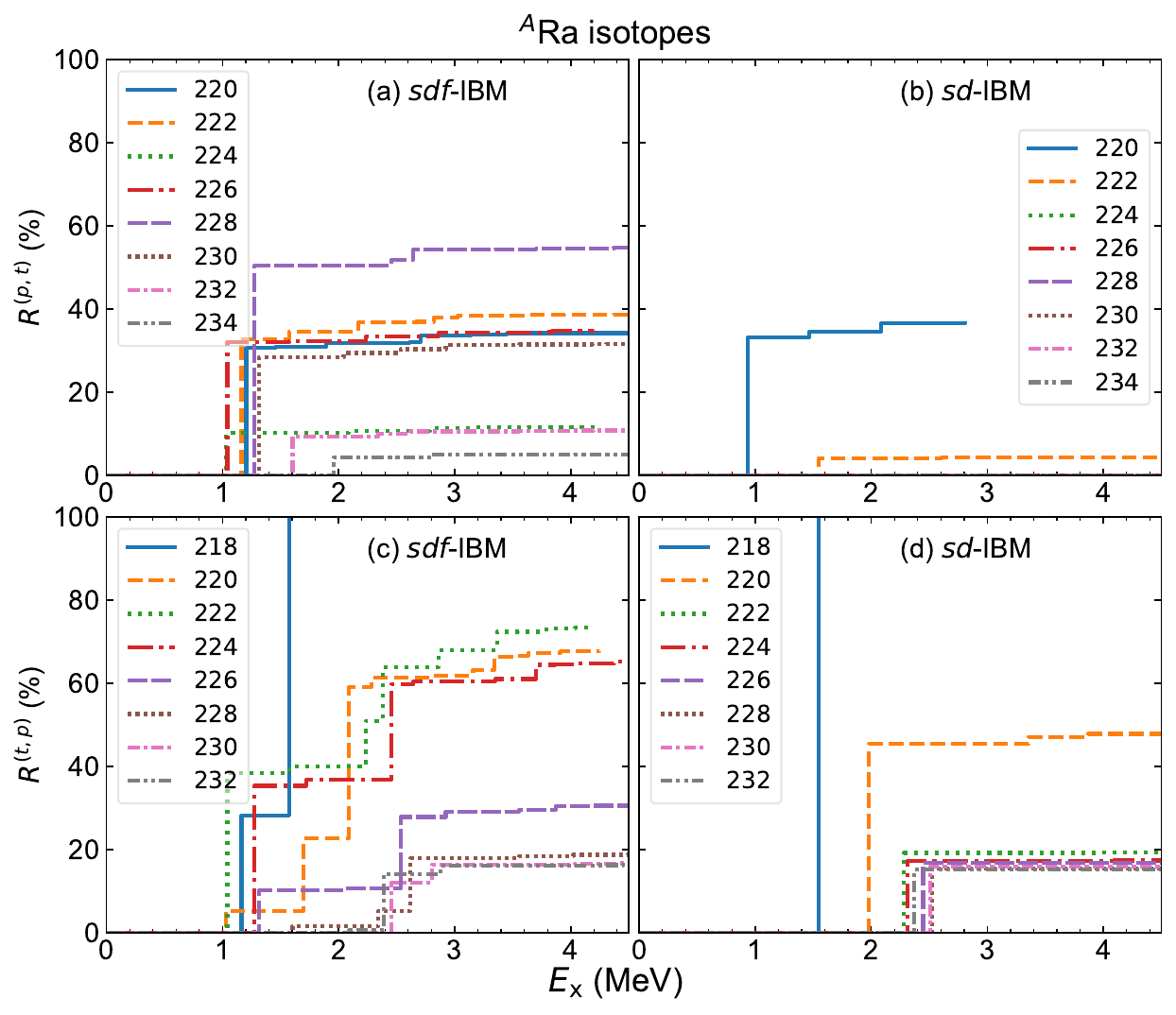}
\caption{Running sums,
$\spt$ and $\stp$,
for $\ptra {A+2} {A}$
and $\tpra {A} {A+2}$ reactions
with $A=218-232$, respectively.
Results from the $sdf$-IBM [(a) and (c)]
and $sd$-IBM [(b) and (d)]
are compared.
}
\label{fig:sum-ra}
\end{center}
\end{figure*}

%-----------------------------------------------------------
%
%   runsum, Th
%
%-----------------------------------------------------------
\begin{figure*}[htb!]
\begin{center}
\includegraphics[width=.7\linewidth]{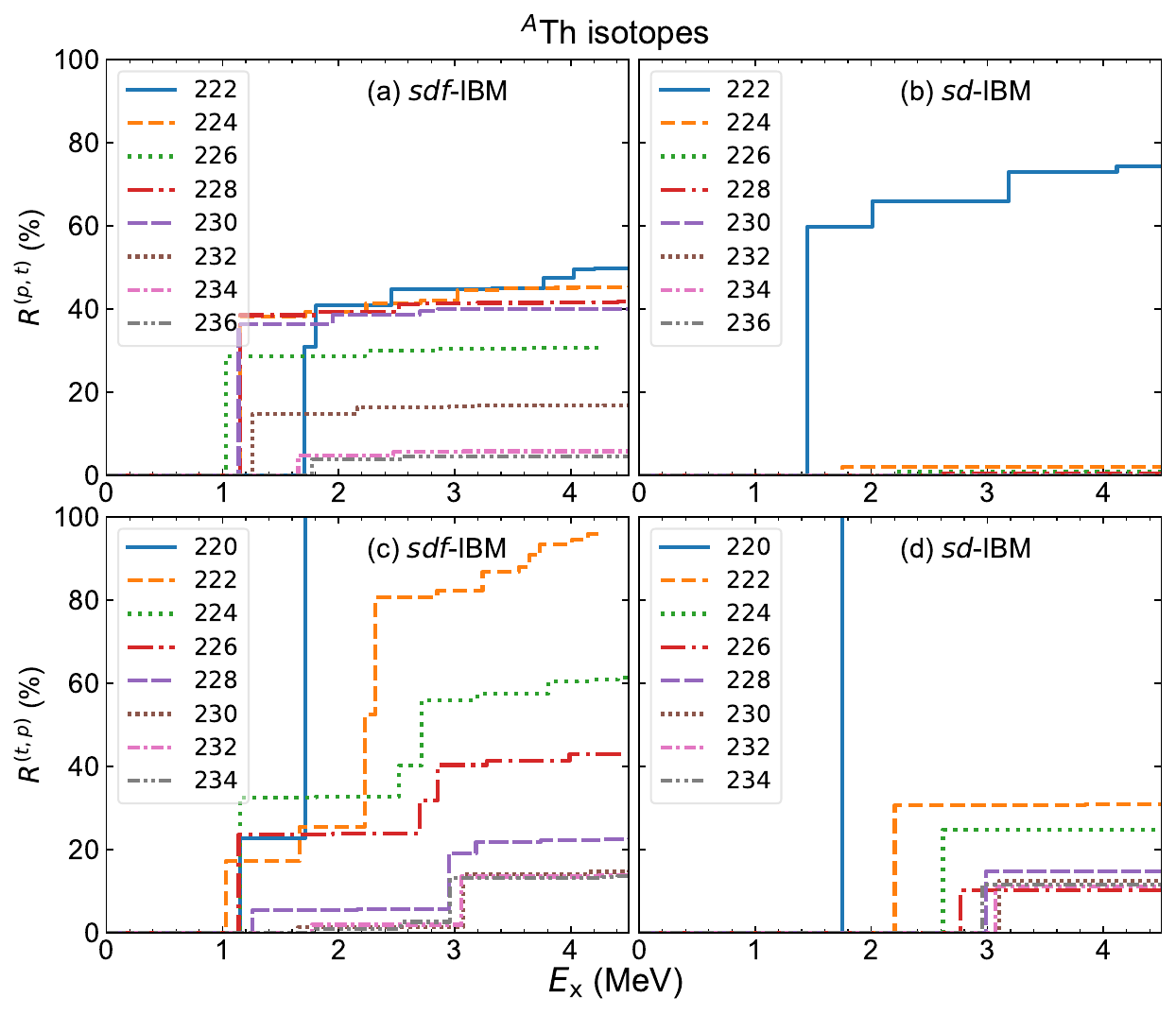}
\caption{As in Fig.~\ref{fig:sum-ra},
but for the $\ptth {A+2} {A}$
and $\tpth {A} {A+2}$ reactions
with $A=220-234$.}
\label{fig:sum-th}
\end{center}
\end{figure*}

\section{Results\label{sec:results}}

\subsection{Low-energy structure}

Figures~\ref{fig:pos} and \ref{fig:neg}
present, respectively, the energy spectra of
even-spin positive-parity
yrast states in the ground-state
($K^{\pi}=0^+_1$) band,
obtained from the $sdf$-IBM and $sd$-IBM,
and the those of odd-spin negative-parity
yrast states, which are considered
to be the members of the $K^{\pi}=0^-_1$
band.
The $sdf$-IBM results for the excitation
energies for these yrast states
are also given in Fig.~4 of
Ref.~\cite{nomura2020oct}.
The differences from the previous calculation
of Ref.~\cite{nomura2020oct} are that,
due to the neglect of the
$\hat L_f\cdot\hat L_f$ term in the
present calculation,
the $K^{\pi}=0^+_1$ band is here
stretched, while $K^{\pi}=0^-_1$ band levels
are systematically lowered;
for example, the $2^+_1$ energy levels
are raised by $\approx10-30$ \%,
and the $1^-_1$ bandhead of the $K=0^-_1$
band is lowered by $\approx 50-150$ keV;
the differences are most significant
for nuclei with $N\approx136$.
These changes, however, do not alter
major conclusions on the
properties of these yrast states.

From Fig.~\ref{fig:pos},
in general, the $sd$-IBM considerably
underestimates the experimental data,
while the $sdf$-IBM energy spectra
are higher than those from the
$sd$-IBM and are more consistent
with the data.
Pronounced differences
between the two IBM predictions
are apparent near $N=132$.
This implies that
the octupole degree of freedom
plays a part in the positive-parity
yrast states within the
mapped $sdf$-IBM.
The negative-parity
energy levels, shown in Fig.~\ref{fig:neg},
exhibit a parabolic behavior as functions
of $N$, being minimal at $N\approx136$.
The Gogny-HFB SCMF calculations
\cite{nomura2020oct}
suggested for many of the Ra and Th
nuclei an octupole-deformed
ground state at
$(\beta_{20},\beta_{30})\approx(0.14,0.11)$,
$(0.16,0.13)$, $(0.18,0.14)$, $(0.20,0.12)$,
and
$(0.22,0.09)$
for $^{220-228}$Ra, respectively,
and at
$(\beta_{20},\beta_{30})\approx(0.14,0.11)$,
$(0.16,0.13)$, $(0.20,0.16)$,
and
$(0.22,0.16)$
for $^{222-228}$Th, respectively.
One can see that the largest
octupole deformation
is suggested at $N\approx136$,
i.e., $^{224}$Ra in
the Ra chain and $^{226,228}$Th
in the Th chain.
The $K=0^{\pm}_1$
bands obtained for these nuclei
were shown to form an
approximate alternating parity
band \cite{nomura2020oct},
while for $N>138$ the predicted
$K=0^-_1$ band increases
in energy and is decoupled from
the $K=0^+_1$ one.
This structural change can be
a signature of the phase transition
from the stable octupole deformed shape to
the octupole-soft shape;
for further details,
see Refs.~\cite{nomura2014,nomura2020oct}.

The octupole degree of freedom has
significant influences
on the nonyrast states,
as illustrated in Fig.~\ref{fig:ny}.
The $sd$-IBM considerably overestimates
the observed $0^+_2$,
$2^+_2$, and $4^+_2$ energy levels,
which exhibit
an inverse parabola.
In the $sdf$-IBM, however,
these energy levels
instead show a parabolic systematic,
being lowest in energy at $N=136$ (for Ra)
and $N=134$ (for Th);
this systematic is
qualitatively consistent
with the observed one \cite{data},
apart from the fact that
the lowest-energy
$0^+_2$ levels are experimentally
found at $N=138$ (Ra) or 140 (Th).
Most significant quantitative
differences between the $sdf$-IBM
and $sd$-IBM predictions
are observed in isotopes with
$N\leqslant 142$,
in which octupole correlations
are expected to be particularly enhanced.
For $N>142$, there is
no notable difference in the calculated
$0^+_2$, $2^+_2$, and $4^+_2$
excitation energies
between the two boson models,
indicating that the octupole bosons
play a less significant part
in these nuclei.

Specifically
for $^{222}$Ra, $^{224}$Ra, $^{226}$Th,
and $^{228}$Th,
the mapped $sdf$-IBM
reproduces the observed nonyrast levels
rather well.
This is in a marked contrast with
the previous mapped $sdf$-IBM
results for the rare-earth nuclei
\cite{nomura2026trf-1},
which significantly overestimated
the experimental $0^+_2$ energy levels.
In the mapped IBM calculations
in general, the overestimates of the
$0^+_2$ energy levels are a consequence
of the unexpectedly large quadrupole-quadrupole
strength.
This is largely accounted for
by incorporating mixing of normal and
intruder configurations in the IBM
\cite{duval1981,nomura2012sc,nomura2016sc}.
It has been pointed out
that the configuration mixing of different
intrinsic shapes and shape coexistence
are crucial for interpreting the
low-energy excited $0^+_2$ states
in the rare-earth nuclei
\cite{garrett2009,tsunoda2023,nomura2025bb-2}.
In the actinide nuclei,
the octupole degree of freedom may seem
to play a more important role
to construct the $0^+_2$ states
than in the case of the rare-earth nuclei.
However, the possibility that
certain shape configuration mixing
occurs in the actinide nuclei
should not be ruled out,
given the considerable overestimates
of the $0^+_2$ levels
in $^{226,228}$Ra and $^{230-234}$Th.

In addition,
the $0^+_2$, $2^+_2$, and $4^+_2$
states can be considered to be members
of the $K^{\pi}=0^+_2$ band.
Some traces of such
$K^{\pi}=0^+_2$ bands are indeed
experimentally suggested
in several actinide nuclei \cite{data}.
As one can see in Fig.~\ref{fig:ny},
the present $sdf$-IBM calculation
suggests that these energy levels lie quite close in energy,
more or less consistently with experiment.
In some nuclei, however,
the $sdf$-IBM does not give
correct order of these energy levels,
whereas the $sd$-IBM
significantly overestimates these levels
but reproduce their order.
These predicted nonyrast states in
the $sdf$-IBM also exhibit
highly complicated $E2$ transition patterns,
and it is therefore not straightforward
to establish a clear band structure
of the excited $K^{\pi}=0^+$
bands within the $sdf$-IBM.

The nature of the $0^+$ states
can be inferred from the calculated
expectation values,
$\braket{0^+_n|\hat n_{f}|0^+_n}\equiv \nf {0} {+} {n}$,
in the corresponding $sdf$-IBM
wave functions,
which provide approximate
$f$-boson content in a given state.
Figure~\ref{fig:nf} displays
the $\nf 0 + n$ values for those
$0^+$ states in Ra and Th nuclei
obtained in the $sdf$-IBM
with the corresponding
excitation energies, $\ezero$, below 6 MeV.
Notably, the $0^+_1$ ground state of
the nuclei $^{220-228}$Ra
and $^{222-228}$Th, in particular,
contains rather large amounts of the
$f$-boson components,
$\nf 0 + 1 \approx 1$,
compared with
the previous mapped $sdf$-IBM
results for Gd, Sm, and Nd isotopes
\cite{nomura2026trf-1},
$\nf 0 + 1 \approx 0.5$.
The result shown in Fig.~\ref{fig:nf}
confirms the importance of
the octupole correlations
to construct the
$0^+_1$ ground state
in those actinide nuclei near $N=134$,
and suggests that the octupole effects
in the actinide region are
more significant than in
the rare-earth region within
the mapped IBM.
For the $0^+_2$ state,
for almost all nuclei studied here,
the values $\nf 0 + 2 \approx 2$ are
obtained, indicating the
two-octupole-boson structure.
A similar conclusion about the $0^+_2$
states was reached for the
rare-earth nuclei
\cite{nomura2015,nomura2026trf-1}.

In deformed rare-earth nuclei,
such as some Gd ones,
a number of low-energy
excited $0^+$ states
are observed experimentally below 3 MeV,
and were considered to arise
from the mixture
of the octupole-boson configurations
in the corresponding $0^+$
wave functions in the $sdf$-IBM
\cite{zamfir2002}.
The situation may be similar
to the actinide region.
Indeed, Fig.~\ref{fig:nf}
suggests that in most nuclei
many $0^+$ states
are obtained at higher energies
above $\approx 3$ MeV, and are
composed of approximately
two or more $f$ bosons.
It should be also noted that
the relevant experimental
information of the nonyrast $0^+$
states, which would support
the above finding,
is rather scarce
in the actinide region.

$E3$ transition properties could be
another observable signature of
the possible double-octupole-phonon
nature of the excited $0^+$ states.
Figure~\ref{fig:rme} gives the
predicted $E3$ reduced matrix elements
between the $3^-_1$ state and lowest
three $0^+$ states,
$|\orme 1|$, $|\orme 2|$, and $|\orme 3|$.
The $\hat T(E3)$ transition operator
is given as $\hat T(E3)=e_3\hat Q_{3}$,
with $e_3$ and $\hat Q_3$ being
the boson effective charge and
octupole operator of
\eqref{eq:q3}, respectively.
By convention,
the same value of the parameters
$\chi_3$ as that used in
the Hamiltonian is employed.
In Ref.~\cite{nomura2020oct},
the $e_3$ charges that
depend on the boson number and
on the $\beta_{20}$ and
$\beta_{30}$ deformations
were considered [see Eq.~(15) of \cite{nomura2020oct}]
in order to incorporate
the octupole effects for those nuclei
near $N=134$ and to better reproduce
the empirical inverse parabolic
systematic of the $\beonp 3 1 0 1$ values.
These $e_3$ values
are adopted in the present work,
which are shown in Fig.~3
of Ref.~\cite{nomura2020oct}.
From Fig.~\ref{fig:rme} one sees
that the calculated
$|\orme 2|$ matrix elements are larger
than the $|\orme 1|$ ones for the
nuclei with $N=132-140$.
This indicates a strong
coupling between
the $0^+_2$ state and
octupole state $3^-_1$.
Note that the predicted
$|\orme 1|$ values are overall
consistent with the experimental
counterparts
\cite{butler2020a,gaffney2013,
wollersheim93,mcgowan1974,mcgowan1993}.

\subsection{Two-neutron transfers}

To study octupole effects
on two-nucleon transfers,
the following quantities
are considered:
\begin{align}
\label{eq:mpts}
& {\mpts 1 n}
=\braket{N-2,0^+_n|\hat A(\Omega_{\nu},N_{\nu})\tilde s|N,0^+_1}\\
\label{eq:mptf}
& {\mptf 1 n}
=\braket{N-2,0^+_n|\hat n_f \tilde s|N,0^+_1}\\
\label{eq:mtps}
& {\mtps 1 n}
=\braket{N+2,0^+_n|s^{\+}\hat A(\Omega_{\nu},N_{\nu})|N,0^+_1}\\
\label{eq:mtpf}
& {\mtpf 1 n}
= \braket{N+2,0^+_n|s^{\+} \hat n_f|N,0^+_1}
\; .
\end{align}
The matrix element in the
first (third) of the above
equations corresponds to the first
term in Eq.~\eqref{eq:pt} [\eqref{eq:tp}],
while the one in the second (fourth)
of the above equations to
the second term of
\eqref{eq:pt} [\eqref{eq:tp}].
In the following, the terms
$\hat A(\Omega_{\nu},N_{\nu})s(^{\+})$
and $s(^{\+})\hat n_f$ are simply referred to as
``$s$-boson'' and ``$f$-boson''
terms, respectively,
and the matrix elements given
in \eqref{eq:mpts}--\eqref{eq:mtpf}
represent their contributions
to the transfer intensities.
Figures~\ref{fig:oct-trf-ra}
and \ref{fig:oct-trf-th} show the calculated
values of the above matrix elements
for the $\ptra {A+2} {A}$
and $\tpra {A} {A+2}$ (with $A=218-234$)
and for the $\ptth {A+2} {A}$
and $\tpth {A} {A+2}$ (with $A=220-236$)
transfer reactions, respectively.

It is seen from Fig.~\ref{fig:oct-trf-ra}
that, for each of the transfer
reactions of Ra,
the $s$-boson contributions
to the ground-state-to-ground-state
($\bej 0 1 0 1$) transfers,
i.e., $\mpts 1 1$ and $\mtps 1 1$,
are dominant over
the ground-state-to-excited-$0^+$-state
($\bej 0 1 0 {\rm ex}$)
ones,
$\mpts 1 {n}$ and $\mtps 1 {n}$
with $n>1$,
in the considered energy
range $\ezero<5$ MeV.
For each transfer reaction,
the $\mpts 1 n$ and $\mtps 1 n$
values become smaller
for higher $\ezero$.
An exception is
the $\tpra {218} {220}$ transfer,
for which the predicted matrix element
$\mtps 1 3$ is particularly large,
being comparable to the $\mtps 1 1$ value.
Both the $\mpts 1 1$ and
$\mtps 1 1$ matrix elements
become monoton larger
for those transfer reactions
of heavier Ra isotopes.
For the $\bej 0 1 0 1$
transfers, the $s$-boson
contributions,
$\mpts 1 1$ and $\mtps 1 1$,
are also considerably larger in
magnitude than those
of the $f$-boson term,
$\mptf 1 1$ and $\mtpf 1 1$.
In some cases, such as
$\ptra {226} {224}$ and $\tpra {224} {226}$,
for which octupole correlations are
particularly strong in the
initial and final nuclei,
the $f$-boson contributions,
represented by the
$\mptf 1 1$ and $\mtpf 1 1$
matrix elements, are also large.
The octupole effects
on the $\bej 0 1 0 1$ $\pt$ and $\tp$
intensities are shown to become minor
for the transfer reactions of
those nuclei with $N$
that are away from $N\approx 134$.

One can observe
from Fig.~\ref{fig:oct-trf-ra}
that the octupole degree of freedom
plays particularly important roles in
the $\bej 0 1 0 2$ strengths.
The $\mptf 1 2$ and $\mtpf 1 2$
values are calculated to be,
in almost all cases, of the
same order of magnitude as
or even larger in magnitude
than the $s$-boson contributions,
$\mpts 1 2$ and $\mtps 1 2$.
For the transfer strengths
to higher-lying nonyrast $0^+$ states,
$\mptf 1 {n}$ and $\mtpf 1 {n}$
are of the same order of magnitude as
$\mpts 1 {n}$ and $\mtps 1 {n}$,
but all these quantities become negligibly smaller
for larger $\ezero$.

Similar results
are obtained for Th isotopes, and are given
in Fig.~\ref{fig:oct-trf-th}.
Apart from some quantitative differences
between the results for the Ra and Th
isotopic chains, one can thus draw similar
conclusions concerning the roles played
by the $f$-boson term and overall systematic
of the transfer intensities,
both as functions of $N$ along
the isotopic chain and of $\ezero$
for a given transfer reaction.

The total transfer intensities,
$\sigpt {1} {n}$
\eqref{eq:sigpt} and $\sigtp {1} {n}$
\eqref{eq:sigtp},
are expected to show characteristic
behaviors as functions of $N$,
and can be thus identified as
a signature of possible QPTs.
Experimental observables that
can be compared with the predicted
transfer intensities are the
cross sections measured at
particular scattering angles.
For the considered Ra and Th isotopes,
however, experimental $\pt$
and $\tp$ data,
which could be used to constrain the $c_1$
and $c_2$ parameters
in the transfer operators,
are rather scarce.
For this reason,
the constant values,
$c_1=0.1$ and $c_2=1$,
are employed for all the transfer
reactions throughout.
In the previous mapped
$sdf$-IBM study of the two-neutron
transfers in rare-earth nuclei
\cite{nomura2026trf-1},
the boson-number and
deformation dependent $c_2$
parameter was proposed.
In the phenomenological $spdf$-IBM
calculations \cite{spieker2013,spieker2018}
constant parameters
for the term involving
negative-parity bosons
in the transfer operators
were considered.

The $\sigpt {1} {1,2}$
and $\sigtp {1} {1,2}$
intensities calculated by
the mapped $sdf$-IBM are depicted
in Figs.~\ref{fig:int-ra} and \ref{fig:int-th}.
One can observe, in Figs.~\ref{fig:int-ra}(a)
and \ref{fig:int-ra}(b),
that the $\sigpt 1 1$ and $\sigtp 1 1$ values
for the Ra isotopes
resulting from the mapped $sdf$-IBM
decrease from
$N=136$ to 138 and from
$N=134$ to 136, respectively.
The $sd$-IBM counterparts
exhibit less significant
variation with $N$.
From a group-theoretical
point of view,
in the limit of
large boson number ($\nb\to\infty$) in
the $sd$-IBM, the matrix element
$|\braket{\nb+1,0^+_1|s^{\dagger}|\nb,0^+_1}|^2$
is proportional to
${\nb+1}$ and
${(\nb+1)(2\nb+3)/3(2\nb+1)}$
in the U(5) and SU(3) dynamical
symmetry limits,
respectively \cite{arima1977trf}.
Along the U(5)-SU(3) phase-transition
chain, such as those observed
in rare-earth nuclei,
the $\mpts 1 1$
value is thus expected to
increase in nearly spherical
region, decrease suddenly
near the critical point, $N=90$,
and start to increase for the
deformed region, $N>90$.
Since the present IBM
involves $f$ bosons,
the comparison with the
symmetry argument within the
$sd$-IBM should be considered to
be approximate.
Nevertheless,
the $\sigpt 1 1$ and $\sigtp 1 1$
values resulting from the
mapped $sdf$-IBM do exhibit certain
traces of possible shape phase
transitions in actinides,
to which octupole
correlations should make
substantial contributions.

The $\sigpt 1 2$ and $\sigtp 1 2$
transfer intensities,
which are shown
in Figs.~\ref{fig:int-ra}(c)
and \ref{fig:int-ra}(d),
also exhibit strong $N$ dependence
in the mapped $sdf$-IBM.
In particular, they become
maximal at $N=140$ [for $\pt$]
and $N=134$ [for $\tp$],
reflecting, to a certain extent,
the lowering pattern of the
energy levels of the
$0^+_2$ states,
which are shown to be made
mainly of the two-$f$-boson
configurations (see Fig.~\ref{fig:nf}).
In the $sd$-IBM with $\nb\to\infty$
\cite{arima1977trf},
the $\bej 0 1 0 2$ $\pt$ transfer
is not allowed in both
the U(5) and SU(3) limits;
the mapped $sd$-IBM values in
Fig.~\ref{fig:int-ra}(c)
follow this in $N>134$,
whereas the $sdf$-IBM gives
finite values.
The $\bej 0 1 0 2$ $\tp$ transfer
is allowed in the SU(3) limit,
but is not in the U(5) limit;
both boson models
in the present study provide
nonzero $\sigtp 1 2$ intensities.
The previous IBM mapping calculations
for the rare-earth Gd, Sm, and Nd isotopes
\cite{nomura2026trf-1}
gave similar $\sigpt 1 2$ and $\sigtp 1 2$
systematic.

The calculated $\sigpt 1 {1,2}$ and
$\sigtp 1 {1,2}$ for the
Th isotopes are shown in Fig.~\ref{fig:int-th}.
The $sdf$-IBM values of
$\sigpt 1 1$ and $\sigtp 1 1$
only monotonically
increase with $N$, hence they do not
appear to exhibit a QPT-like behavior
as the one observed in the Ra isotopes.
This can be partly accounted for
by the adopted value of the
parameter, $c_2=1$, which may not be
appropriate for Th.
In fact it is confirmed that
increasing slightly the $c_2$ value to,
e.g., $c_2\approx1.5$,
leads to a similar
$\sigpt 1 1$ and $\sigtp 1 1$
systematic as that obtained for
the Ra isotopes.
In Figs.~\ref{fig:int-th}(a)
and \ref{fig:int-th}(c),
experimental data
for the $\ptth {230} {228}$
and $\ptth {232} {230}$ transfer
reactions, adopted from
Ref.~\cite{wirth2004},
are shown.
These data are measured cross sections
at the laboratory angle $7.5^{\circ}$,
estimated from the distorted-wave
Born approximation (DWBA) analysis \cite{wirth2004}.
The data
suggest an increase from $N=140$ to 142.
The $\pt$ intensities produced
by the $sdf$-IBM exhibit a similar trend.
As in the case of the Ra chain,
the $sd$-IBM suggests a monotonic,
weak $N$ dependence of the $\sigpt 1 1$
and $\sigtp 1 1$ values, except for the
lightest Th nuclei.

The predicted
$\sigpt 1 2$ and $\sigtp 1 2$
intensities for Th in the $sdf$-IBM
exhibit peaks at $N=140$ and $N=134$,
respectively, and these behaviors are
similar to those obtained
for the Ra isotopes.
The experimental cross sections
for the $\pt$ transfers
\cite{wirth2004} indicates an increasing
pattern from $N=140$ to 142.
The present $sdf$-IBM values of
$\sigpt 1 2$, however,
decrease from $N=140$ to 146.

To infer gross features of the
$\pt$ and $\tp$ strengths to
excited $0^+$ states,
running sums of the relative
transfer intensities
\begin{eqnarray}
\label{eq:sum}
R^{\alpha}
=\sum_{n=2}
\frac{I^{\alpha}(\bej 0 1 0 {n})}
{I^{\alpha}(\bej 0 1 0 1)}
\; ,
\end{eqnarray}
with $\alpha=\pt$ or $\tp$,
normalized with
respect to the transfer
strength to the ground state,
are considered.
Figures~\ref{fig:sum-ra} and \ref{fig:sum-th}
give the corresponding results
for Ra and Th isotopes, respectively.
It is seen from Fig.~\ref{fig:sum-ra}(a)
that, for all Ra isotopes,
the $\bej 0 1 0 2$ transfers
dominate the $\pt$ strengths
to the excited $0^+$ states
in the considered energy range
of $\ezero\leqslant4.5$ MeV
in the $sdf$-IBM:
The sums, $\spt$,
converge at the $0^+_2$ excitation
energies, which are typically
$1-1.5$ MeV.
For the
$\ptra {232} {230}$ and
$\ptra {234} {232}$ $\pt$
reactions,
the running sums
amount to $\spt\approx 10$ \%,
and thus transfer strengths to the excited
$0^+$ states are of minor importance.
The $sdf$-IBM suggests that
the $\spt$ sums for the
$\ptra {228} {226}$ and
$\ptra {224} {222}$
reactions converge to particularly
high ($\approx 50$ \%)
and low ($\approx 10$ \%) values,
respectively,
as compared with those
for the $\pt$
reactions of neighboring isotopes.
The peculiar tendencies
are expected from the
behaviors of the $\sigpt 1 2$ values,
being small at $N=136$
and maximal at $N=140$
[see, Fig.~\ref{fig:int-ra}(c)],
which reflect, to a large
extent, the octupole
correlation effects.
In the $sd$-IBM, there is essentially
no contribution from
the $\pt$ transfers to excited $0^+$
states, except for the
$\ptra {220} {218}$ and
$\ptra {222} {220}$
reactions.

As one sees in Fig.~\ref{fig:sum-ra}(c),
the running $\tp$ strengths
continue to increase with $\ezero$.
In the majority of cases,
there appear two major rises of the
running sums: one at the $0^+_2$
excitation energies
and the other at
$\ezero\approx 2-3$ MeV.
The sums appear to converge
at $\ezero>3$ MeV.
Only for the $\tpra {218} {220}$
reaction does the sum exceed
the $\sigtp 1 1$
intensity at the excitation energy
of the $0^+_3$ state in $^{220}$Ra.
The $\stp$ values for
$\tpra {220} {222}$,
$\tpra {222} {224}$, and
$\tpra {224} {226}$ reactions
reach more than 60 \%
at $\ezero>2$ MeV.
For those $\tp$ transfers
of the Ra nuclei with mass $A\geqslant 228$,
contributions of the $\bej 0 1 0 {\rm ex}$
transfers become less significant
in comparison with the
$\bej 0 1 0 1$ ones.
These results are expected from
the systematic behaviors of the
$\sigtp 1 {1,2}$ intensities,
shown in Figs.~\ref{fig:int-ra}(b)
and \ref{fig:int-ra}(d).
In the $sd$-IBM,
the $\tp$ running sums, $\stp$,
shown in Fig.~\ref{fig:sum-ra}(d),
are almost purely accounted for
by the $\bej 0 1 0 2$
transfer, while there is
no contribution from
the transfer strengths
to higher-lying $0^+$ states.

Figure~\ref{fig:sum-th}
gives the running sums calculated
for the Th isotopes.
Similarly to the Ra case,
the $sdf$-IBM suggests that
the $\bej 0 1 0 2$ transfers
are most important
among the
$\bej 0 1 0 {\rm ex}$
strengths, with the running sums
being converged approximately at the
$0^+_2$ energies,
and that contributions from
the transfer strengths
to higher-lying
$0^+$ states are only incremental.
The $\spt$ values become smaller
for transfer reactions
of heavier Th isotopes.
However, the converged $\spt$ values
for the $\ptth {228} {226}$
and $\ptth {230} {228}$
reactions are larger than
that of the $\ptth {226} {224}$ one.
This irregularity arises from the large
$\sigpt 1 2$ values calculated
for the $\pt$ transfers of $^{228}$Th
and $^{230}$Th, as compared with
those of other nuclei;
see Fig.~\ref{fig:int-th}(c).

One can compare the mapped $sdf$-IBM
running sums
with the experimental counterparts
for the $\ptth {230} {228}$ \cite{levon2013}
and $\ptth {232} {230}$ \cite{levon2009}
reactions.
In Refs.~\cite{levon2009,levon2013},
phenomenological $spdf$-IBM and
quasiparticle-phonon model (QPM)
calculations were also performed.
It was shown in Ref.~\cite{levon2009}
that the measured running sum for the
relative $\ptth {232} {230}$
transfer intensity
exhibited a significant rise
at the $0^+_2$ excitation energy
in $^{230}$Th to amount to $\spt\approx30$ \%,
and a second rise in
the range $\ezero\approx 1.5-2.5$,
increasing from
30 \% ($\ezero\approx1.5$ MeV)
to 60 \% ($\ezero\approx 2.5$ MeV).
Both the $spdf$-IBM and QPM
reproduced this trend.
The $\pt$ relative
running sum predicted by
the $sdf$-IBM for $\ptth {232} {230}$
exhibits only a single rise
at the $0^+_2$ state in $^{230}$Th,
and converges to $\spt\approx$ 40 \%.
The experimental running sum for the
$\ptth {230} {228}$ transfer was
shown \cite{levon2013} to be rather stable
at $\approx 20$ \%
from $\ezero\approx 0.6$ MeV,
corresponding to the $0^+_2$ excitation
energy, to $\ezero\approx1.5$ MeV,
but exhibited a second,
steady rise for $\ezero>1.5$ MeV.
The QPM calculation in
Ref.~\cite{levon2013} qualitatively
reproduced this trend.
In the present $sdf$-IBM,
the relative running sum
for the $\ptth {230} {228}$ reaction
reaches $\stp\approx 20$ \%,
but exhibits only a single rise
at the $0^+_2$ excitation energy.

The systematic of the $\tp$ running sums
in Th isotopes, shown in Fig.~\ref{fig:sum-th}(c),
appears to be similar to that
of Ra isotopes,
that is, the converged values of
the running sum
become smaller for the $\tp$
transfers of heavier isotopes,
and a second rise of the sums
appears within the energy range of
$\ezero\approx2.5-3$ MeV.
The results obtained from the
$sd$-IBM, shown in Figs.~\ref{fig:sum-th}(b),
and \ref{fig:sum-th}(d)
are also quite similar
to those for the Ra isotopes.

\section{Conclusions\label{sec:summary}}

Octupole correlation effects on
the excited $0^+$ states and two-neutron
transfer intensities in actinide nuclei
have been investigated in the
framework of the $sdf$-IBM,
using the interaction strength parameters
determined by means of the microscopic
SCMF method.
The calculated energy levels
of the $0^+_2$ states in
the $sdf$-IBM for most of the
considered nuclei
were shown to be significantly
lower and closer to experimental
values than those of the $sd$-IBM.
The $sdf$-IBM also turned out to
give an improved description
of the positive-parity
even-spin yrast states in
the ground-state bands.
The $sdf$-IBM produces a large
number of excited $0^+$ states,
most of which are composed
of the configurations
containing more than one $f$ boson.

Octupole correlations have been
shown to have significant influences on
two-neutron transfer intensities
in actinide nuclei.
For the $\bej 0 1 0 {2}$
$\pt$ and $\tp$ transfer
intensities,
contributions from the
$f$-boson term in the
transfer operators turned out
to be larger than those from
the leading-order, $s$-boson term,
particularly for those nuclei
with $N=132-138$,
in which enhanced
octupole collectivity is expected.
Furthermore,
the mapped $sdf$-IBM predicts
some characteristic behaviors of the
$\pt$ and $\tp$ intensities
for the $\bej 0 1 0 {1,2}$
transitions as functions of $N$
along each isotopic chain,
which can be regarded as
signatures of the possible
shape phase transitions.
The $sd$-IBM, however, gives
monotonic evolution with $N$ of
these transfer intensities.

The octupole degree of freedom appears
to be relevant to the
mapped IBM description of the low-lying states
of both parity and two-nucleon
transfers in the actinide region.
The results shed lights upon the
interpretation of the low-energy $0^+_2$
states and corresponding bands
observed in light actinides
in terms of the double-octupole-phonon
structure.
The overestimates of the $0^+_2$ energy
levels, which still persist in some Ra and Th nuclei with
$N\geqslant 138$, may indicate yet
another correlations that are
missing in the present version
of the theoretical framework,
e.g., configuration mixing.
These points will be addressed
as a future study and reported elsewhere.

\acknowledgements
This work has been supported by JSPS
KAKENHI Grant No. JP25K07293.

\bibliography{refs}

\end{document}